# How Can Bar Robots Enhance the Well-being of Guests?

Oliver Bendel [1] and Lea K. Peier [2]

[1] *School of Business FHNW, Bahnhofstrasse 6, 5210 Windisch, Switzerland*
[2] *xappido ag, Länggasse 3, 6208 Oberkirch, Switzerland*

#### Abstract

This paper addresses the question of how bar robots can contribute to the well-being of guests. It first develops the basics of service robots and social robots. It gives a brief overview of which gastronomy robots are on the market. It then presents examples of bar robots and describes two models used in Switzerland. A research project at the School of Business FHNW collected empirical data on them, which is used for this article. The authors then discuss how the robots could be improved to increase the well-being of customers and guests and better address their individual wishes and requirements. Artificial intelligence can play an important role in this. Finally, ethical and social problems in the use of bar robots are discussed and possible solutions are suggested to counter these.

## 1. Introduction

In many science-fiction stories and films, a robot serves the guests in a bar, café, or restaurant. It is designed to be human-like or machine-like and is often as quick-witted and witty as its human counterparts. Sometimes it plays a central role, sometimes a marginal one. Think of the android Arthur, the bartender in the 2016 film "Passengers", or the short story "The Crystal Crypt" by Philip K. Dick [1]. The visitor receives important information or is well entertained – or is simply served.

In reality, more and more service robots and social robots are spreading in cafés, restaurants, shopping malls, and airports [2]. Bar robots are behind the counter and prepare coffee or cocktails. Transport and serving robots help to fill and clear buffets and serve guests, with the first prototypes already being created in the 1990s [3]. These are joined by other machines with specific tasks, such as security and cleaning. It seems worthwhile to study gastronomy robots scientifically.

This paper first elaborates on the basics of service robots and social robots. It gives a concise overview of which gastronomy robots are on the market. Then it presents examples of bar robots – to reduce complexity, this will be the sole focus – and describes two models currently in use. A research project by the authors at the School of Business FHNW, which included a thesis, collected empirical data on them [4]. The paper then discusses how the robots and the settings could be improved to increase the well-being (i.e., satisfaction, comfort, health, and pleasure) of customers and guests, and to better address their individual wishes and requirements. Artificial intelligence (AI) can play an important role in this. Finally, ethical and social problems in the use of bar robots are discussed and solutions are suggested to counter these.

## 2. Basics of Robots in Gastronomy

The following section discusses robots of all kinds in gastronomy and catering. First, features of gastronomy are explained. Second, the terms "service robot" and "social robot" are clarified. Third, examples of robots in gastronomy are given.

## 2.1. Gastronomy Characteristics

Gastronomic establishments offer meals and drinks in a suitable setting, usually in publicly accessible rooms or houses with tables, chairs, and pleasant music. Guests devote themselves to satisfying basic human needs, i.e. satisfying hunger and thirst. Equally important to many, however, is the opportunity to socialize and meet new people. The houses and rooms are centers of attraction, offering security, comfort, and entertainment. Eating and drinking together enables and strengthens social relationships.

The bartender or waitress either plays the role of a conversation partner when there are no, hardly any or not the desired customers on site. Or they mediate in multifaceted relationships. Replacing or supporting a bartender therefore has an impact on the social fabric. Gastronomic establishments can also be part of shopping malls, train stations, and airports.

## 2.2. Types of Robots

Robots can be broadly classified into two types: industrial robots and service robots. Classic industrial robots operate in protected areas of the factory, primarily in production. Modern variants are cobots that work hand in hand with humans. Service robots take over services and provide assistance in human settings. Examples include transportation robots, security robots, and cleaning robots. Some service robots are basically converted industrial robots. For example, one takes a cobot and uses it on a stationary or mobile platform for services.

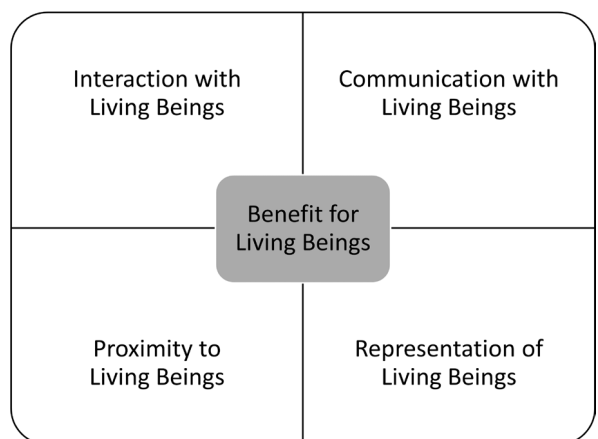

**Figure 1**: The five dimensions of social robots [5]

Social robots are sensorimotor machines created to interact with humans or animals, particularly more sophisticated species [5]. They can be determined through five key aspects. These are: interaction with living beings, communication with living beings, representation of (aspects of or features of) living beings (e.g., they have an animaloid or a humanoid design or natural language abilities), proximity to living beings, and fundamentally, utility or benefit for living beings (see Fig. 1). The assumption is that they are social robots if four or five dimensions are met.

Some social robots are service robots, that is, they handle certain services and provide certain assistance, and conversely, some service robots are also social robots, insofar as they have communication and interaction functions. Typical examples in this intersection are care and therapy robots. A care robot like Lio is a service robot that was created by putting a cobot, i.e. an industrial robot, on a mobile platform and equipping it with certain capabilities. These include social features, so the robot can be considered a social robot.

Precisely these distinctions are essential for gastronomy robots. These can be counted among the service robots. Some are social robots, or at least have social characteristics. This is important in cafes and restaurants where people are social and should feel safe and comfortable. The robots should not be foreign bodies and should approach people in an appropriate manner, talk to them and be considerate. Some models are also industrial robots adapted for the service sector. For example, bar robots in particular are mostly cobots with one or two arms.

## 2.3. Robots in Gastronomy

Gastronomy is a complex area with different tasks and requirements. In a facility there is often a kitchen, a bar with a counter and bar stools, and an area on one level or on several levels with tables and chairs. The office, staff room, and reception area are available to the staff or are used for the initial contact with the customer. Accordingly, quite different robots appear [6].

First of all, there are bar robots [4]. They are mostly cobots that have special tasks and abilities that make them service robots. They are housed in a kiosk (a kind of glass cage, also referred to as encapsulation below) or behind the counter. They prepare coffee, as so-called barista robots [7, 8], or cocktails, as so-called cocktail robots. Other drinks such as juice mixtures are also possible.

They either do this completely on their own or need help with individual steps, such as decorating the cocktails.

There are also transport and serving robots. They support the staff by moving back and forth between the buffet and the kitchen, loaded with empty and full plates and containers [2]. Or they serve guests and assist waiters in bringing food and drinks and clearing dishes from tables [3, 9]. Examples include BellaBot from Pudu Robotics (www.pudurobotics.com), with a cat face on the display and plastic cat ears, Plato from Aldebaran or United Robotics Group (cobiotx.unitedrobotics.group), with a waiter fly under the display, and Lucki from OrionStar (en.orionstar.com). In addition, humanoid models exist. Delivery robots are a special feature in this context – they leave the business to visit customers in the area or in other parts of the city, either remotely or autonomously.

Robots can also be found in the kitchen. They help prepare food or automatically make doughs, sauces, and soups [10]. Machines such as food processors or steamers have, of course, been present in the kitchen for a long time, but robots have only been around for a short time. Here again, robotic arms dominate, similar to how they are used in industry. Cobots and kitchen staff can work safely hand in hand in a confined space and play to their respective strengths.

Last but not least, robots welcome guests to facilities. At the moment, this is mainly present in hotels, where humanoid, animaloid, or other models replace or supplement the staff at the reception. Some of them even resemble dinosaurs [11]. Certainly, more and more restaurants will also make use of this possibility, especially those where one is guided to the table. Furthermore, cleaning robots can be used to clean floors and other surfaces. A distinction must be made between wet and dry cleaning [7]. Last but not least, security robots can control exterior and interior areas [7].

## 3. An Overview of Bar Robots

The following is an overview of bar robots. It is based primarily on a literature review and interviews with experts that took place as part of the thesis [4]. It is not complete and is primarily intended to illustrate their worldwide use and different forms of implementation. It should be noted at the outset that most companies do not manufacture the arms used themselves, but rather purchase them from companies such as KUKA, for example, and integrate them into their systems. One exception is F&P Robotics.

1. Makr Shakr is a robotics company based in Turin, Italy (www.makrshakr.com). Its products include Toni and Toni Compatto. Toni is a two-armed robot that accesses 158 different bottles hanging from the ceiling of the open bar and shakes and stirs the liquids. Toni Compatto is a smaller version. It is equipped with one arm and requires only half the space.
2. MyAppCafé is a company from Marxzell in Germany, whose product range includes the My-App-Café (my-app-cafe.com), a coffee station in a kiosk operated by a robot. The bar is equipped with a robot barista, two fully automated coffee machines, an ice cube machine, two syrup stations, and a foam printer.
3. Crown Digital is a Singapore company (www.crowndigital.io). Its products include Ella, a barista robot [12]. The robot arm is located in a kiosk with glass walls. Ella prepares coffee, which guests order either via the app or directly from the display at the kiosk.
4. Rossum Cafe is a company from Slovakia (rossumcafe.com). Its barista robot is an encapsulated bar. Orders are placed either via the Rossum Cafe device or the app. Different types of coffee and teas can be ordered.
5. Cafe X are kiosks from San Francisco in the USA with a robotic arm (cafexapp.com). They serve hot and cold drinks, but mainly specialty coffees from local roasters in the shortest possible time. The bar robot can access coffee machines and taps.
6. Swiss Smyze from Zurich in Switzerland (smyze.com) is the provider of Robobarista. This consists of a robotic arm and the dispensing, ice, and coffee machines. All components are housed in a 2.20 meter high and 2.20 meter wide kiosk with glass walls.
7. F&P Robotics from Glattbrugg near Zurich in Switzerland uses the P-Rob, its robotic arm, not only for its care robot Lio, but also for Barney (www.fp-robotics.com). Barney can act as a cocktail or barista robot (Barney Bar and Barney Barista). It operates freely behind the counter. The finishing of cocktails has to be performed by humans.
8. FoodTemi is a concept by scientists from Taiwan that stands out from the other examples. "The overall system integrates the Dobot arm, a linear motion robot, and Temi mobile robot. By ordering the drinks via the panel on robots, the Dobot arm and linear motion robot will prepare the beverage. When the meal is ready, FoodTemi will automatically fetch and deliver the meal to the customer." [13]

As has become clear, bar robots are capable of preparing coffee, cocktails, or both. They can perform some operations and steps autonomously. For others, they require human assistance. This also means that they do not displace the bartender in all cases, but stand by him or her and perform tasks together with him or her.

Alongside the drinks or food, the zeitgeist and a lifestyle feeling are served, for example with the offer of local and sustainable roasting companies. An element of spectacle is also important – the components are freely visible behind the counter or clearly visible behind the glass walls. Some companies like Makr Shakr and Cafe X advertise the speed of their robots as if they were cars.

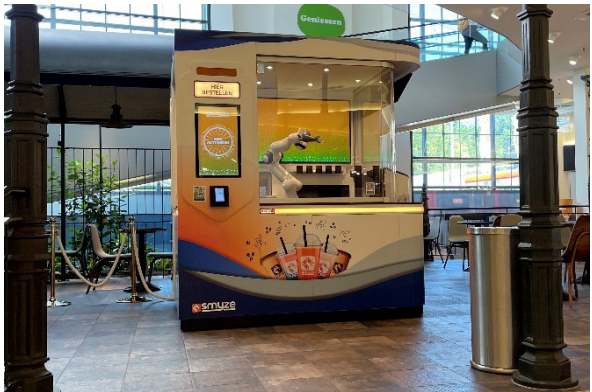

**Figure 2**: Robobarista from Swiss Smyze

Single-arm robots predominate. However, two-armed robots are also available as they are in industry (think of YUMI from ABB). They can solve certain tasks better or differently – for example, they can hold something with one gripper and manipulate something with the other. These models seem more like living beings or like people than one-armed ones and thus raise certain expectations.

## 4. An Investigation of Bar Robots

In this section, Barney Bar and Robobarista are discussed in detail. These two models were chosen because they were available in Switzerland at shopping malls and supermarkets. Not only were expert interviews conducted on the two robots, but they were also observed in action in Swiss cities over several weeks as part of the thesis [4].

First, the two bar robots are described. Basic functions have already been mentioned above; this is now more about the manufacturers and their ambitions with regard to AI. Then the bar robots are assigned to the five dimensions of social robots. This is important because later it must be clarified how the well-being of the guests can be promoted when they are used. The dimensions are also needed for this. The descriptions are taken from the documentation on the robots, the expert interviews with the managers of the companies, and the own observations in the facilities.

### 4.1. Robobarista

The startup, Swiss Smyze AG, was founded in 2020 [4]. The Robobarista (which is not only a barista but also a bartender) was launched in early 2022. The company cooperates with various domestic and foreign suppliers who provide the major components. Swiss Symze AG assembles them into the Robobarista and programs the associated software.

**Table 1**
Description of Robobarista [4]

| Dimension | Specification |
| --- | --- |
| Benefit | The Robobarista provides a chargeable service by preparing and serving coffee and mocktails. In this way, it supports or supplements service staff and serves customers. |
| Interaction | The Robobarista interacts with the user via screens. On the order screen, it takes requests. On other displays, it shows the status of the beverage preparation or pictures of the finished beverages. |
| Communication | The Robobarista can emit sounds and thus communicate with humans. It did not do this during operation. |
| Representation | Actually, the Robobarista should not depict a living being, according to the expert interview. Nevertheless, it can be associated with human arms or an animal. The movement is also reminiscent of living beings. |
| Proximity | The proximity to living beings is not as pronounced as with other bar robots. The Robobarista is located in a kiosk with glass walls, so it is encapsulated. |

The Robobarista can be found in Swiss shopping malls, among other places (see Fig. 2). It consists of a robotic arm and the dispensing, ice, and coffee machines. All components are housed in a kiosk with glass walls. The robot has the overall appearance of a vending machine and, because it is encapsulated, can in principle be set up anywhere where there is a power and water supply and an internet connection.

AI has not yet been fully developed for the Robobarista [4]. As soon as several robots are in use and a lot of data can be generated, AI algorithms will also be applied. This could include, for example, adapting the menu to the weather. Currently, only industry standards, such as collision protection, have been implemented.

## 4.2. Barney Bar

F&P Robotics is an SME and offers care robots and bar robots, as well as cobots for industry [4]. Barney Bar and Barney Barista are sold by F&P Robotics and can be found in Switzerland (see Fig. 3) as well as in various shopping centers in Oman and in a convention center in China.

The P-Rob robot arm is used not only for Lio, but also for Barney. Barney can act as a cocktail or barista robot. It operates freely behind a counter. The finishing of cocktails or mocktails has to be done by humans. So they add mint leaves or lemon slices.

**Table 2**
Description of Barney Bar [4]

| Dimension | Specification |
| --- | --- |
| Benefit | Barney Bar provides a free service by preparing and serving mocktails, i.e. cocktails without alcohol, as a promotion of the bar in question. In this way, it supports or complements service staff and serves customers. |
| Interaction | Barney Bar communicates with people via tablet and screen. The order can be placed via a tablet, and a name can be entered. Via a screen, it shows the guests the current status of the preparation of the drinks. |
| Communication | Communication between Barney and people takes place by telling bartender jokes. This function was not used during operation. |
| Representation | Actually, Barney should not depict a living being, according to the expert interview. Nevertheless, it can be associated with human arms or an animal. The movement is also reminiscent of living beings. |
| Proximity | The proximity to living creatures is pronounced. Guests can even touch the robot arm behind the counter. |

F&P Robotics is trying to use AI algorithms to make the robot more interesting. The neural network in Barney Bar – which is the focus here – refers to recognizing and grasping different containers. It is trained with different objects on a regular basis. As a result, it can play with glasses or cups, by pretending to drop something. This is also a popular trick among human bartenders and helps to entertain customers.

Barney Barista from F&P Robotics could additionally be viewed after the expert interview at the headquarters. It is located in a training room where it is being tested for various functionalities. The bar has an open design. One could touch the coffee machine, the open bar, or even the robot arm. The design of the bar is kept simple and traditional and is mainly made of wood.

Barney Barista has a screen that can display the menu, various videos, or the status of the order. Customers place their orders via a tablet integrated into the bar. They choose a drink, enter their name, as in Barney Bar, and pay for the drink at the payment terminal. On the screen behind the robotic arm, a display appears that resembles a messaging app on a smartphone. This informs the customer which drink is currently being prepared.

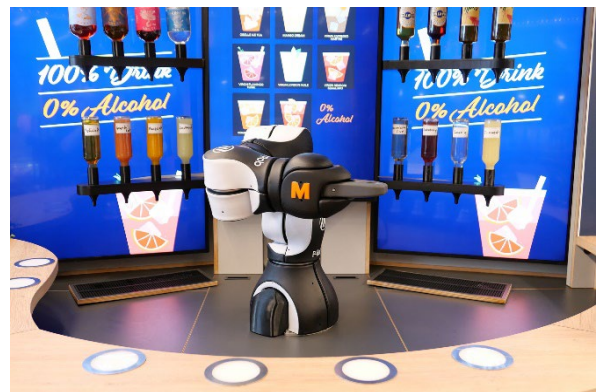

**Figure 3**: Barney Bar from F&P Robotics

## 5. Use of Bar Robots in Switzerland

Robobarista and Barney Bar were in operation in several cities in Switzerland in 2022 [4]. The companies placed them in shopping malls and supermarkets. Gastronomic "islands" were created there, with the aim of offering customers an attraction and providing them with drinks (and thus keeping them in the store longer).

The co-author systematically observed both robots as part of her thesis in the summer of 2022. Tally sheets were kept on usage and impact, and individual customer interviews were conducted where permitted.

## 5.1. Guest Monitoring and Surveys

The monitoring at Swiss shopping centers and supermarkets involved 620 people (Barney Bar: 253, Robobarista: 367). A total of 90 customers took a closer look at the two robots, 72 percent of them Barney Bar and 28 percent Robobarista. 50 people ordered a drink, that is, more than half of those who were at the bar [4].

The majority reacted positively and seemed to be interested in technical and procedural aspects. Certainly, the possibility of receiving a free drink (which was the case with Barney) was also attractive. Last but not least, the guests could feel like pioneers who had first contact with a bar robot and could tell their family and friends about it. Some of them took photos with their cell phones, some selfies.

From the surveys of consumers and potential guests, it is clear that a bar robot is eye-catching. Its technical possibilities and implementation fascinate some people. However, the lack of personal contact tends to be unsettling and become a hindrance. It also runs counter to the interviews with the gastronomy businesses that emphasized how guests appreciate a personal, honest, and sincere exchange with human staff. The need to communicate with a service robot or a social robot is obviously not yet there.

## 5.2. Surveys of Gastronomists

The four people interviewed from gastronomy businesses (representatives of a café, a restaurant, a café chain, and a bar) emphasize that they cannot imagine bar robots being used in their businesses [4]. In their opinion and that of the authors, it would also be difficult to set up Barney Bar or Robobarista in restaurants, cafés, or bars due to the space available. As a consequence, sales areas in the form of seats would ultimately have to be taken away.

The café owner surveyed emphasized that it is impossible for a bar robot to make good coffee, as it cannot smell, feel, or see it during preparation. The gastronomy businesses also rated the standardization of drinks and, above all, the portioning of ingredients as a cause for concern. They obviously perceive themselves as establishments that can respond well to individual wishes and sensitivities and thus increase the well-being of customers.

## 5.3. Specific Observations

The expert discussions and observations have produced or reinforced some further findings. The bar robots cannot respond to specific customer wishes and needs. Nor can a robot muster the passion with which employees in the gastronomy perform their job, along with the associated emotions that are impressed upon the clientele. It should also not be forgotten that attractive waiters and waitresses are often hired who appeal to the clientele on a sexual level, however this may be judged. Cobots cannot offer this either. Last but not least, it can be assumed that the novelty effect only generates interest at the beginning. This could not be verified by the observations. However, individual statements by interviewees indicate that this is the case.

## 6. Improvements for Well-being

This section proposes improvements to bar robots and their use to provide guests and customers with a pleasant experience and contribute to satisfying their needs, giving them comfort, and strengthening their health, i.e., contributing to their well-being. The idea that a gastronomic establishment is a social environment is maintained. Conclusions from the expert interviews and observations are incorporated, as well as the authors' own experiences – both have worked as service staff in gastronomy during their studies.

The model already presented is again used for structuring. The five dimensions are considered separately below. However, it is clear that the interplay of characteristics and capabilities can reinforce an effect.

## 6.1. Interaction

Individual needs of the guests could be taken into account during the interaction. For example, after identification – via facial recognition or a login – the bar robot could know that it is dealing with an allergy sufferer, diabetic, vegetarian, vegan, Muslim, or Jew. Accordingly, it could avoid or add certain ingredients when preparing the drinks. This contributes to the satisfaction and health of the customer.

An individually created profile would go beyond this. The user could indicate his or her preferences in any respect, such as taste, or ecological and ethical aspects. For example, he or she could prefer organic ingredients from local cultivation, of course in coordination with the capabilities of the machine and the components.

Emotion recognition using facial recognition or voice recognition would also be an option. So if the guest seems sad, the robot could say encouraging words or offer a free drink. Conversely, if he or she is happy, the robot could raise the mood. Emotion recognition can also be used to study the guest's reactions and optimize the bar robot and setting.

### 6.2. Communication

With regard to communication, it was found that existing functions of the bar robots were not used at all or only to a limited extent. In the case of Barney Bar, the manager's explanation was that the ambient noise was too loud, and the microphones were therefore switched off. The same problem would be present in restaurants and cafés. However, the guest can sit close to the robot at the counter, so communication could succeed there after all.

The bartender jokes that Barney has already mastered could be activated in a suitable environment. They contribute to the pleasure of the visitors. The name entered at Barney Bar or Barney Barista, for example, also plays a role. As the manager explained in the expert interview, the robot could say at the appropriate time: "Hi Joe, here comes your espresso" (name was changed).

ChatGPT based on GPT-3.5 has achieved great attention in 2022 [14]. One can integrate such a system into a bar robot and thereby make it as conversational as desired. This gives the user the dialogue and entertainment they want. However, the machine counterpart does not really understand what he or she is saying, nor does it have any actual interest in the guest.

Multilingualism is an enhancement that can be implemented using translation programs, which in turn can be based on AI. This means that, unlike the average bartender, the robot can address people from all cultural and linguistic backgrounds and provide them with a pleasant experience. The customer feels not only understood in the literal sense, but also valued.

### 6.3. Representation

Most of the bar robots, including the two discussed in detail, have one arm. They look a bit like living beings, which is strengthened by the purposeful and sure movement. Some have two arms, reinforcing the impression and making one think of a human or non-human primate. This could contribute to the guests' comfort. Two arms allow for different and simultaneous activities. Three or more arms could be used to magnify its role as a spectacle but would create a high level of complexity and weaken the humanoid impression.

Lio, the care robot based on the same robotic arm as Barney, mostly has two magnetic eyes in operation, which have no function but an effect: Patients perceive it as a snake, a bird, or even a human [15]. Such additions could also be applied to bar robots, both one-armed and two-armed. The concept is called robot enhancement [16]. One could also use other accessories, although these work better for humanoid robots.

### 6.4. Proximity

The place of the bar robots in Switzerland at the moment is the shopping mall and the supermarket. They form gastronomic "islands" there. Conviviality can hardly arise, even if the observations showed that some guests got into conversation with each other. Bar robots could be integrated into cafés and restaurants. However, based on the expert interviews and the authors' own experience, it can be said that this would be difficult in many establishments because there is not enough space. In contrast, such installations could certainly be made in modern, larger establishments, and they could be planned into new buildings.

The Robobarista seems to allow less social proximity than Barney Bar or Barney Barista. It is basically a robot in the guise of a vending machine. Barney, on the other hand, is as close as a bartender and can even be touched. If it moves quickly, one can feel a breeze. It is no longer an automaton, not even in appearance. This, in turn, could contribute to the guests' comfort.

### 6.5. Benefit

The benefit of bar robots is that they provide a service by preparing and serving cocktails or coffee. In this way, they support or complement service staff and serve customers. With the help of the improvements mentioned above, the benefit expands. The bar robots could become bartenders, entertaining customers, conversing with them, and knowing and taking into account their needs.

## 7. Ethical and Social Discussion

The following is a brief ethical and social discussion. This is arranged according to problem areas. The five dimensions of social robots are still involved but are not in the foreground.

The section clearly shows that the well-being that can be mediated by bar robots can be compromised by several factors. This makes it all the more important to take the ethical and social challenges seriously.

### 7.1. Overrated Conversation Skills

If the bar robot has outstanding communication capabilities, for example based on GPT-3 or GPT-4, the guests have the impression that someone really understands them, takes them seriously, and values them. This is reinforced by an animaloid or humanoid impression. It can, however, be associated with the risk of deception and fraud [17].

This could be counteracted by recurring statements by the bar robot that it is just a machine. This has been implemented in chatbots and voice assistants, for example [18]. Even if the effect is little researched (perhaps the user ultimately does not care that he or she is dealing with a machine), transparency is at least established.

### 7.2. One-sided Relationships

The guest, who is actually looking for social proximity, including to the bartender, is fobbed off with a one-sided relationship but could be satisfied with this [5]. In extreme cases, he or she reduces genuine human contact. Social places, which are the cafés or the restaurants, become antisocial places, machines become things for lonely people to project their loneliness upon. Here, too, one can speak of deception and fraud [17].

This could also be counteracted by recurring statements by the bar robot that it is just a machine. Other V-effects (effects of alienation and disillusionment according to the concept of the German dramatist Bertolt Brecht) are also possible [18]. For example, the bar robot might make noises and sounds that are not attributed to a human or that usually disrupt a relationship. It should be added that some bartenders are exposed to sexual harassment by patrons. This situation can be improved by bar robots when they replace or assist their human counterparts, with the help of monitoring functions in dangerous hotspots. In addition, the machine can become a conversation piece. Then it helps to form real social relationships.

### 7.3. Personal Data Drain

If the bar robot has a camera, if it performs facial recognition and associated emotion recognition, there is a risk that personal data will be processed and passed on. In this context, the situation of consuming alcohol can be problematic. In many cases, the guest will not have informational autonomy, i.e., will not be able to view, check, change, and delete the data.

It would be desirable to create appropriate options and return informational autonomy to the user. This could be handled via the robot's display, the robot's natural language capabilities, or via the account that one sets up. In addition, existing image-processing technology should be used as sparingly as possible and only after consent has been obtained.

### 7.4. Replacement of Employees

Bar robots can replace employees in whole or in part. In addition, they can take away activities that one enjoyed doing and that were of high quality, such as preparing coffee with the right aroma, stirring and shaking cocktails, or making coffee, along with demonstrating tricks. Such activities may even constitute the bartender's self-image. If he or she is relegated to decorating the cocktail, it could have a significant impact on job satisfaction.

It is important to make sure that bar robots support rather than replace bartenders. When providing support, care must be taken to ensure that central and meaningful activities are not lost for the employee. He or she could, however, exploit the user's wonder at the robot and use it to his or her advantage. He or she would appear, at least initially, as someone who has mastered and owns an innovation. However, this variant of the novelty effect will certainly wear off.

## 7.5. Accidents and Damages

In a kiosk, accidents involving guests are unlikely to happen. With robotic arms that move freely in the space behind the counter, on the other hand, this is possible. It is true that cobots are made to be in close proximity to humans. But in an open-access environment, inexperienced users can become a problem. In principle, they can be injured by the robot if they move too quickly, touch it, or get in its way.

The solution here lies in design or technology. One can shield the free arm from the user at certain points without using a full glass and metal housing. Or one can improve the cobot, which is already good at handling close proximity to people, even further for the specific situation. For example, one could reduce the speed at which drinks are prepared, which would admittedly run counter to its role as a spectacle.

## 7.6. Mistakes and Misconceptions

If the bar robot takes into account guests' world views, diets, and illnesses by avoiding ingredients when preparing drinks, there are opportunities and risks in this. Needs can be satisfied and health strengthened. However, they can also be impaired if mistakes are made. Questions then also arise about responsibility and liability. The robot itself cannot bear any responsibility.

It is important that the user can create and control his or her own profile. The robot's processes must be checked regularly, as must the physical sources it accesses, such as the bottles or tubes on the ceiling. Ultimately, errors cannot be avoided, but they could even be fewer than with service staff, who may not even know and be able to assess diets and illnesses. In this sense, the robot could contribute to standardization and quality improvement and ultimately to the health of the guests.

## 8. Summary and Outlook

This paper was dedicated to the question of how bar robots can increase the well-being of guests. First, it outlined the basics of service robots and social robots. It found that some service robots are social robots and many social robots are service robots. It is also true that some service robots are based on the technology of industrial robots. The model with the five dimensions of social robots was used for classification and determination.

Gastronomy robots can be counted among service robots, although they are based on industrial robots and may be social robots. The paper gave a concise overview of what exists in terms of gastronomy robots. It then presented examples of bar robots, describing Barney Bar and Robobarista and their use in Switzerland. The authors then discussed how robots and settings could be improved to enhance the well-being of customers and guests and better address their individual needs and requirements. Different possibilities of AI were mentioned, such as communication possibilities based on GPT-3 or GTP-4, facial recognition, and emotion recognition. Finally, ethical and social problems in the use of bar robots were discussed and solutions were suggested to counter these.

Bar robots can contribute to the well-being and health of guests, serving them drinks they like and providing a special experience. At the moment, there are reservations among restaurateurs and guests. These must be taken seriously, and manufacturers must respond to them. The aforementioned improvements can transform the bar robot, to some extent, into a bartender and focus on the needs of the customer. In the process, useful functions are possible that were not or hardly available before. In several cases, these have to do with artificial intelligence. However, the bar robot can hardly replace the operator in social matters. It is quite conceivable that it will establish itself as a support for the bartender – along the lines of a classic cobot.

## 9. References

[1] P. K. Dick, Sämtliche 118 SF-Geschichten: Der Philip-K.-Dick-Companion, Zweitausendeins, Leipzig, 2018.

[2] O. Bendel, Serviceroboter aus Sicht der Ethik, in: M. Lindenau, M. Meier Kressig (Eds.), Schöne neue Welt? Zwischen technischen Möglichkeiten und ethischen Herausforderungen, Vadian Lectures, vol. 6, transcript, Bielefeld, 2020, pp. 57–76.

[3] B. A. Maxwell, L. A. Meeden, N. A. Addo, The Robot Waiter Who Remembers You, in: AAAI Technical Report WS-99-15. URL: https://www.aaai.org/Papers/Workshops/1999/WS-99-15/WS99-15-001.pdf.

[4] L. Peier, Ein Serviceroboter in einem Café hinter der Bar, Bachelor Thesis at the School

of Business FHNW, School of Business FHNW, Olten, 2022.

[5] O. Bendel (Ed.), Soziale Roboter: Technikwissenschaftliche, wirtschaftswissenschaftliche, philosophische, psychologische und soziologische Grundlagen, Springer Gabler, Wiesbaden, 2021.

[6] J. M. Garcia-Haro, E. D. Oña, J. Hernandez-Vicen, S. Martinez, C. Balaguer, Service Robots in Catering Applications: A Review and Future Challenges, Electronics, 10 (1), 47 (2021).

[7] O. Bendel, 450 Keywords Digitalisierung, 2nd ed., Springer Gabler, Wiesbaden, 2022.

[8] B. Lemm, Roboter in der Gastronomie: Das Café der Zukunft, 2021. URL: https://www.gastroinfoportal.de/news/gastroinfoportal-beverage-heissgetraenke/das-cafe-der-zukunft/.

[9] F. Seyitoğlu, S. Ivanov, Understanding the robotic restaurant experience: a multiple case study, Journal of Tourism Futures, 8 (1), 18 November 2020. URL: https://www.emerald.com/insight/content/doi/10.1108/JTF-04-2020-0070/full/html.

[10] A. Zoran, Cooking With Computers: The Vision of Digital Gastronomy [Point of View], in: Proceedings of the IEEE, vol. 107, no. 8, 2019, pp. 1467–1473.

[11] O. B. Waxman, Dinosaurs Have Come Back from Extinction as Hotel Receptionists In Japan, Time, 16 July 2015. URL: https://time.com/3960847/dinosaur-robot-receptionist-japan/.

[12] Z. Salim, S'pore's first robot barista ELLA to be deployed at 30 MRT stations, Vulcan Post, 19 August 2021. URL: https://vulcanpost.com/757320/robot-barista-ella-to-be-deployed-at-30-mrt-stations/.

[13] C.-F. Hung, Y. Lin, H.-J. Ciou, W.-Y. Wang, H.-H. Chiang, FoodTemi: The AI-Oriented Catering Service Robot, in: 2021 IEEE International Conference on Consumer Electronics-Taiwan (ICCE-TW), 2021, pp. 1–2.

[14] C. Stokel-Walker, AI bot ChatGPT writes smart essays – should professors worry? Nature, News Explainer, 9 December 2022. URL: https://www.nature.com/articles/d41586-022-04397-7.

[15] O. Bendel, A. Gasser, J. Siebenmann, Co-Robots as Care Robots, in: Proceedings of the AAAI 2020 Spring Symposium "Applied AI in Healthcare: Safety, Community, and the Environment", 2020. URL: https://arxiv.org/abs/2004.04374.

[16] O. Bendel, Möglichkeiten und Herausforderungen des Robot Enhancement, in: S. Schleidgen, O. Friedrich, J. Seifert (Eds.), Mensch-Maschine-Interaktion – Konzeptionelle, soziale und ethische Implikationen neuer Mensch-Technik-Verhältnisse, Mentis, Münster, 2022, pp. 267–283.

[17] H. Schulze, O. Bendel, M. Schubert et al., Soziale Roboter, Empathie und Emotionen. Zenodo, Bern, 2021. URL: https://zenodo.org/record/5554564.

[18] O. Bendel, Das GOODBOT-Projekt, in: O. Bendel (Ed.), Handbuch Maschinenethik (Springer Reference Geisteswissenschaften), Springer VS, Wiesbaden, 2019.